\documentclass[twocolumn,showpacs,preprintnumbers,amsmath,amssymb]{revtex4}
\usepackage{graphicx}
\usepackage{dcolumn}
\usepackage{bm}

\begin{document}

\preprint{APS/123-QED}
\title{Observation of Lasing Mediated by Collective Atomic Recoil}

\author{D. Kruse}
\author{C. von Cube}
\author{C. Zimmermann}
\author{Ph.W. Courteille}
\affiliation{Physikalisches Institut, Eberhard-Karls-Universit\"at
T\"ubingen,
\\Auf der Morgenstelle 14, D-72076 T\"ubingen, Germany}

\date{\today}

\begin{abstract}
We observe the buildup of a frequency-shifted reverse light field
in a unidirectionally pumped high-$Q$ optical ring cavity serving
as a dipole trap for cold atoms. This effect is enhanced and a
steady state is reached, if via an optical molasses an additional
friction force is applied to the atoms. We observe the
displacement of the atoms accelerated by momentum transfer in the
backscattering process and interpret our observations in terms of
the collective atomic recoil laser. Numerical simulations are in
good agreement with the experimental results.
\end{abstract}


\maketitle

Since the first observation of photonic recoil \cite{Frisch33}, a
large variety of techniques has been developed to exploit the
mechanical forces of light for optical cooling and trapping. The
inverse process, i.e. the impact of the atomic motion on light
fields, has recently received special attention with the
theoretical prediction \cite{Guo92} and the observation
\cite{Courtois94,Kruse03} of the phenomenon of recoil-induced
resonances (RIR) and with the proposal for a collective atomic
recoil laser (CARL) \cite{Bonifacio94}. In RIR experiments a pump
and a weak probe laser beam having different frequencies give rise
to a moving standing light wave at the intersection region. Atoms
moving synchronously to this wave undergo two-photon Raman
transitions between momentum states. This gives rise to a density
grating at the intersection region, which copropagates with the
standing wave. The density modulation corresponds to a classical
bunching of atoms in the optical potential valleys formed by the
standing wave. The Raman scattering process can equivalently be
described as stimulated Rayleigh scattering of photons off the
density grating \cite{Courtois94} from one beam into the other. If
the probe laser is red-detuned from the pump, it receives a net
gain. The probe gain increases the amplitude of the standing wave.
This leads to stronger atomic bunching and, in turn, strengthens
the Rayleigh scattering in an avalanche process. In RIR
experiments the relative gain is typically small, so that runaway
amplification of the probe is not observed. Consequently, the
positive feedback described above is neglected in common RIR
theories.

CARL which shares the same gain mechanism as RIR \cite{Berman99}
has been formulated as a transient process \cite{Bonifacio94}, the
signature of which is an exponential growth of a \textit{seeded}
probe field oriented reversely to a strong pump interacting with
an active medium. On the other hand, atomic bunching and probe
gain can also arise spontaneously from fluctuations with no seed
field applied \cite{Bonifacio95}. The underlying runaway
amplification mechanism is particularly strong, if the reverse
probe field is recycled by a ring cavity. So far, attempts to
observe CARL action have only been undertaken in hot atomic vapors
\cite{Lippi96,Hemmer96}. They have led to the identification of a
reverse field with some of the expected characteristics. However,
the gain observed in the reverse field can have other sources
\cite{Brown97}, which are not necessarily related to atomic
recoil, so that the unambiguous experimental proof of the CARL
effect is still owing.

The CARL effect should emerge most clearly in cold atomic clouds
in a collisionless environment \cite{Berman99}. Furthermore, large
detunings of the lasers far outside the Doppler-broadened profiles
of the atomic resonances are preferable. In this regime effects
from atomic polarization gratings, which are not based on density
variations \cite{Brown97}, are avoided. Finally, to emphasize the
role of the exponential gain responsible for selfbunching, i.e.
spontaneous formation and growth of a density grating, it is
desirable not to seed the probe. The observation of a probe beam
is then a clear indication for CARL in contrast to RIR, where the
avalanche effect is nonexistent. In this paper we present
experiments which fulfill the above requirements.

  \begin{figure}[ht]\centerline{\scalebox{0.55}{\includegraphics{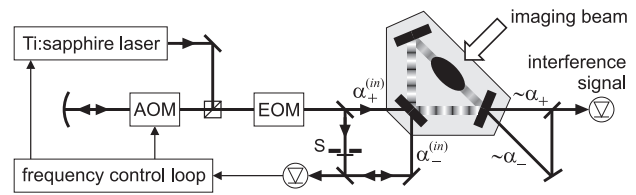}}}
  \caption{Scheme of the experimental setup. A titanium-sapphire
  laser is locked to one of the two counterpropagating modes ($\alpha_+$) of
  a ring cavity. The beam $\alpha_-^{(in)}$ can be switched off by means of a
  mechanical shutter (S). The atomic cloud is located in the free-space
  waist of the cavity mode. We observe the evolution of the interference
  signal between the two light fields leaking through one of the
  cavity mirrors and the spatial evolution of the atoms via absorption
  imaging.}
  \end{figure}

Our high-$Q$ ring cavity consists of one plane and two curved
mirrors. It is $8.5$~cm long, has a beam waist of $w_0=130~\mu$m
and an amplitude decay rate of $\kappa=2\pi\times22~$kHz. The
cavity is pumped by a titanium-sapphire laser. For the sake of
definiteness, we describe in the following the cavity modes by
their field amplitudes scaled to the field per photon
\cite{Gangl00}. The mode $\alpha_+$ is continuously pumped at a
rate $\eta_+=\sqrt{\delta \kappa}~\alpha_+^{(in)}$, where $\delta$
is the free spectral range of the cavity and $\alpha_+^{(in)}$ the
field amplitude of the incoupled laser beam. The titanium-sapphire
laser is stabilized to this mode by the Pound-Drever-Hall method.
The counterpropagating mode is labelled by $\alpha_-$. We call
$\alpha_+$ the pump and $\alpha_-$ the probe mode. The details of
the apparatus have been published in Ref.~\cite{Kruse03}. We load
$^{85}$Rb atoms from a magneto-optical trap (MOT) into a
TEM$_{00}$ mode of the ring cavity. With typically $N=10^6$ atoms
trapped, we achieve atom peak densities of more than
$2\times10^{11}$~cm$^{-3}$ at temperatures of several $100~\mu$K
corresponding roughly to $1/5$ of the potential depth. The
temperature of the atomic cloud and its density distribution are
monitored by time-of-flight absorption imaging. We measure the
intracavity light power via the fields leaking through one of the
cavity mirrors (intensity transmission $T=1.8\times10^{-6}$, see
Fig.~1). The outcoupled light power is related to the intracavity
power by $P_{\pm}^{(out)}=T P_{\pm}^{(cav)}=T\hbar \omega
\delta~|\alpha_{\pm}|^2$. The phase dynamics of the two
counterpropagating cavity modes is monitored as the beat signal
between the two outcoupled beams. Any frequency difference between
pump and probe, $\Delta\omega\equiv\omega_+-\omega_-$, i.e.
propagation of the standing wave nodes inside the ring cavity, is
translated into an amplitude variation of the observed
interference signal $P_{beat}=T \hbar \omega
\delta~|\alpha_++\alpha_-|^2$.

  \begin{figure}[ht]\centerline{\scalebox{0.43}{\includegraphics{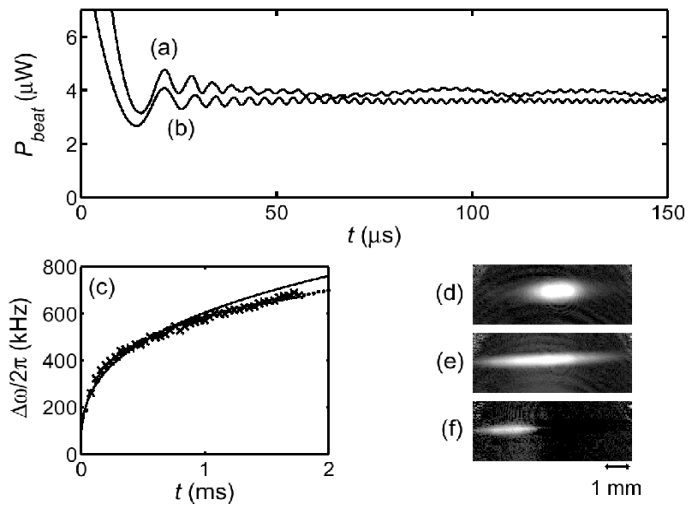}}}
  \caption{\textbf{Curve (a)}: Recorded time evolution of the observed
  beat signal between the two cavity modes with $N=10^6$ and $P_{\pm}^{(cav)}=2~$W.
  At time $t=0$ the pumping of the probe $\alpha_-$ has been interrupted.
  \textbf{Curve (b)}: Numerical simulation according to Eq.~(1) with the temperature
  adjusted to $200~\mu$K. \textbf{(c)}: The symbols (x) trace the evolution
  of the beat frequency after switch-off. The dotted line is based on a
  numerical simulation. The solid line is obtained from Eq.~(3) with the
  assumption that the fraction of atoms participating in the coherent dynamics
  is $1/10$ to account for imperfect bunching. \textbf{(d)}: Absorption images
  of a cloud of $6\times 10^6$ atoms recorded for high cavity finesse at $0~$ms
  and \textbf{(e)} $6~$ms after switching off the probe beam pumping. All images
  are taken after a $1~$ms free expansion time. \textbf{(f)}: This image is obtained
  by subtracting from image \textbf{(e)} an absorption image taken with low cavity
  finesse $6~$ms after switch-off. The intracavity power has been adjusted to the
  same value as in the high finesse case.}
  \end{figure}

In order to observe a CARL signal, we load atoms from the MOT into
the symmetrically pumped ring cavity standing wave
($\lambda=797.0~$nm). The detuning from the nearest atomic
transition ($D$1-line at $794.8~$nm) is $1~$THz. After $30~$ms
trapping time one beam ($\eta_-$) is switched off via a mechanical
shutter. With no atoms the interference signal $P_{beat}$ then
drops to $T \hbar \omega \delta~|\alpha_+|^2$ within a time of
$10~\mu$s, limited by the finite shutter closing time. An observed
residual low-frequency fluctuation of $P_{beat}$ is assigned to
scattering at the imperfect mirror surfaces. In contrast, if atoms
are loaded into the ring cavity standing wave, oscillations appear
on the interference signal shortly after the switch-off, as shown
in Fig.~2(a). Their amplitude is rapidly damped, however they are
still discernible after more than $1.5~$ms, which exceeds the
cavity decay time $\tau=(2\kappa)^{-1}$ by more than a factor of
$600$. We verified that the oscillations do not occur for low
cavity finesse, which is realized by rotating the polarization of
the incoupled lasers from $s$ to $p$ polarization \cite{Kruse03},
thereby reducing the finesse from $80000$ to $2500$. Figure~2(c)
shows the frequency of the interference signal as a function of
time. From the observations we can deduce 1.~that the probe mode
$\alpha_-$ is fed with light in the presence of atoms, thus
leading to a standing wave superposed to the pump mode $\alpha_+$.
2.~Having verified that the oscillations are solely due to a
relative phase shift of the cavity modes (no oscillations are
detected in either mode), we know that the detuning between probe
and pump increases in time, and therefore the standing wave is
displaced and accelerated by the presence of atoms. 3.~With time
the contrast of the standing wave reduces to zero, i.e. after a
fast initial decay, the probe fades out on a time scale of
$1.5~$ms. 4.~The atoms are dragged by the moving standing wave.
The frequency shift of the backscattered light field by up to
$1~$MHz corresponds to an atomic velocity of $40~$cm/s. At
interaction times of a few ms, the atomic motion should lead to a
detectable displacement. To verify this we have taken
time-of-flight absorption images of the atomic cloud at various
times after one-sided switch-off, shown in Fig.~2(d) and (e). We
observe that, even though a major fraction of the cloud keeps
staying at the waist of the laser beam, its center-of-mass is
shifted along the propagation direction of the standing wave. In
the case of low cavity finesse, the displacement is almost zero.

The observations can be understood in terms of lasing mediated by
collective atomic recoil, despite the differences between our
system and the original CARL proposal \cite{Bonifacio94}. The
original idea is based on injecting a homogeneously distributed
monokinetic atomic beam counterpropagating to the pump and a seed
for the probe, which predetermines the probe frequency. In this
system, the signature for CARL action is a blue-detuned amplified
probe pulse followed by a highly irregular evolution of the
system. In contrast, in our system the probe frequency and the
atomic momentum distribution evolve in a selfconsistent manner.
Initially the atoms are highly bunched at the antinodes of the
symmetrically pumped standing wave cavity field. After switching
off $\eta_-$, those atoms which take part in the CARL process
transfer photons from the pump into the probe and by recoil,
acquire momentum, $mv$, in direction of the pump. The frequency
difference between probe and pump corresponds to twice the Doppler
shift, $\Delta\omega=2kv$. Thus, in agreement with our
observations, an increasingly red-detuned probe beam is expected,
whose intensity eventually diminishes, because its frequency
drifts out of the cavity resonance by an amount corresponding to
$\Delta\omega$.

Being focussed on studies of transient phenomena, the original
CARL model does not consider relaxation for the translational
degrees of freedom. The long-term dynamics predicted by this model
yields atomic acceleration without any bounds \cite{Perrin02b}. To
obtain steady state operation, the addition of dissipation to the
momentum equations has been suggested \cite{Bonifacio96,Perrin01}.
We translate this idea to experiment by introducing a friction
force by means of a superimposed optical molasses: In a second set
of experiments, we load the atoms from the MOT directly into the
unidirectionally pumped running wave dipole trap ($\eta_- = 0$).
In contrast to the first set of experiments, we now start from a
basically homogeneous atomic density distribution, so that the
appearance of light in the probe mode constitutes an evidence for
selfbunching. After $40~$ms trapping time, we switch on the
optical molasses. We use the laser beams of the MOT and tune them
$50~$MHz below the cooling transition ($D$2, $F=3 \rightarrow
F'=4$). Fig.~3(a) shows the interference signal observed when the
optical molasses is turned on at time $t=0$. Apparently, strong
oscillations emerge from the noise floor. They quickly reach a
steady frequency between $100$ and $170~$kHz, which corresponds to
an atomic velocity of $7$ to $13~$cm/s, and they persist for times
longer than $100~$ms, mainly limited by the finite size of the
molasses region. Fig.~3(c) demonstrates the dependence of the beat
frequency on the pump rate $\eta_+$.

  \begin{figure}[ht]\centerline{\scalebox{0.43}{\includegraphics{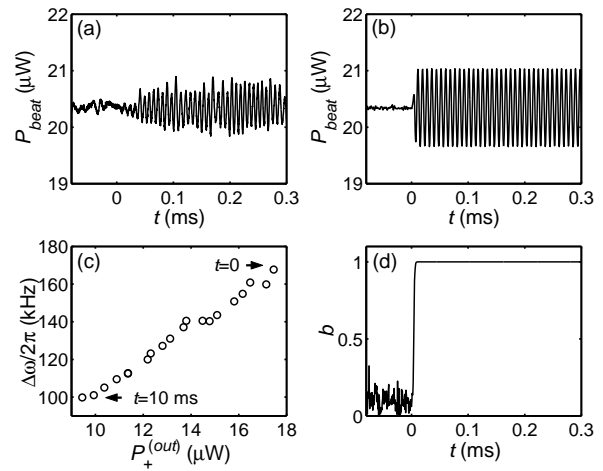}}}
  \caption{\textbf{(a)}: Formation of a moving standing wave upon
  irradiation of molasses beams at time $t=0$. The experimental settings
  are the same as in Fig.~2(a) except for the light power $P_+^{(cav)}=11~$W.
  \textbf{(b)}: Numerical simulation of the beat signal obtained with a
  unidirectionally pumped cavity ($\eta_-=0$) according to Eq.~(1).
  The atom number and the friction coefficient are adjusted to $N=2\times
  10^5$ and $\gamma_{fric}=9\kappa$, respectively. \textbf{(d)}: The
  simulation also yields the bunching parameter $b\equiv N^{-1}\left| \sum_ne^{2ikx_n}\right|$
  defined in Ref.~\cite{Bonifacio94}. The perfect bunching for $t>0$ is an artifact of our
  simplified friction model, which does not account for diffusion heating by the molasses.
  \textbf{(c)}: Impact of the pump rate $\eta_+$ on the steady state
  propagation velocity of the standing wave. The pump rate is ramped down
  within $10$~ms and monitored through the outcoupled laser power $P_+^{(out)}$,
  which decreases from $18$ to $10~\mu$W.}
  \end{figure}

We interpret the experimental observations in the following way:
Any inhomogeneity in the atomic density distribution, e.g. due to
a residual standing wave ratio in the cavity field generated by
back-scattering from the cavity mirrors, gives rise to some amount
of atomic bunching. Without the optical molasses, this bunching is
not sufficient to initiate runaway CARL amplification, mainly
because the atoms are accelerated and redistributed in space,
before a relevant bunching of the atomic density can occur. The
damping force of the molasses now counteracts the acceleration
force and prevents dispersion of the atomic velocities. A
steady-state velocity is reached, when the velocity dependent
damping force balances the CARL acceleration. The molasses thus
acts like a pump recycling the atoms into an equilibrium momentum
state that otherwise would be depleted by CARL acceleration. The
equilibrium defines the frequency of the probe, which therefore is
dependent on the intensity of the pump laser, as is shown in
Fig.~3(c), and the molasses parameters. We found that the
frequency also depends on the atom number.

We model our system consisting of an ensemble of atoms coupled to
two counterpropagating modes of a ring cavity with a set of
differential equations \cite{Bonifacio95,Gangl00,Perrin02}. We
concentrate on the limit of very large laser detuning from the
atomic resonances, where the excited atomic states can be
adiabatically eliminated (Eqs.~(16-18) in Ref.~\cite{Gangl00}).
Adapting this model to our experiment, we have to account for the
fact that one mode, the pump $\alpha_+$, is tightly phase-locked
to the cavity. Any phase variation of $\alpha_+$ is translated by
the servo loop into a phase correction fed back to the incoupled
laser $\alpha_+^{(in)}$, thus suppressing any relative phase
dynamics. Being interested mainly in the situation of one-sided
pumping of the cavity ($\eta_-=0$), we also neglect variations of
$\alpha_+$ due to photon scattering from and into the (almost)
empty probe $\alpha_-$. We then obtain a stationary field
amplitude $\alpha_+=\chi^{-1}~ \eta_+$, where we introduced the
abbreviation $\chi=\kappa+i N U_0-i\Delta_c$. Here $N$ is the atom
number, $U_0$ the one-photon light shift and $\Delta_c$ the
detuning of the laser from the resonance of the empty cavity. The
very large detuning from the atomic resonances allows us to
neglect the radiation pressure exerted by the cavity fields. The
backscattered probe and the motions of the atoms located at the
positions $x_n$ are then described by
\begin{align}
\dot{\alpha}_- & =-\chi\alpha_--\tfrac{iU_0\eta_+}{\chi}
\sum\nolimits_ne^{2ikx_n},\label{Eq01}\\
k\ddot{x}_n & =2\varepsilon iU_{0}\eta_+\left( \tfrac{\alpha_-
}{\chi^{\ast}}e^{-2ikx_n}-\tfrac{\alpha_-^{\ast}}{\chi}e^{2ikx_n}\right)
-\gamma_{fric}k\dot{x}_n,\nonumber
\end{align}
where $k$ is the wavenumber of the field and
$\varepsilon\equiv\hbar k^2/m$ is twice the recoil shift. Note
that we have phenomenologically included a damping term
$\gamma_{fric}kv$ to account for molasses friction. This procedure
has been introduced in Ref.~\cite{Bonifacio96} to model the impact
of atomic collisions. In our experiment, collisions are
negligible.

In general the equations cannot be solved analytically, and we
have to numerically iterate them. For the sake of computational
efficiency we treat groups of $10^4$ atoms as single particles.
Fig.~2(b) and Fig.~3(b) trace the simulated evolution of the
coupled system consistent of the atomic cloud and two optical
modes of the ring cavity as pump and probe. We find good agreement
with the experimental data. The simulations also reveal that part
of the atoms copropagate with the moving standing wave potential,
sitting on its back slope and thus experiencing a constant
accelerating light force. This force comes from the photonic
momentum transfer accompanying the light scattering into the
reverse mode: The atoms behave like surfing on a self-sustained
standing light wave. Without friction the atoms continuously
accelerate, and the standing wave ratio and the atomic bunching
gradually decrease. The inclusion of friction leads to increased
backscattering, deeper potential valleys and stronger atomic
bunching.

Analytic solutions can be derived for the case of perfect atomic
bunching. We may then replace the atomic variables by
center-of-mass coordinates $x\equiv N^{-1}\sum_n x_n$. Inserting
the ansatz $\alpha_-\equiv\beta e^{2ikx}$, with the assumption of
unidirectional pumping and steady state conditions ($\eta_-=0$,
$\dot{\beta}=0$), one obtains
\begin{align}
\beta & =\tfrac{-iNU_0\eta_+}{\chi\left( \chi+2ikv\right) }
,\label{Eq02}\\
k\dot{v} & =\tfrac{4\varepsilon NU_0^2\eta_+^2}{|\chi|^2}
\operatorname{Re}\left(\tfrac{1}{\chi+2ikv}\right)
-\gamma_{fric}kv~.\nonumber
\end{align}
For simplicity we assume $\Delta_{c}=N U_{0}$, which is generally
satisfied when the laser is locked to the cavity, whose resonance
is shifted by the presence of atoms \cite{Elsasser03}. Ignoring
friction ($\gamma _{fric}=0$) and with the initial condition
$v(t=0)=0$, the differential equation in (2) is solved by
\begin{equation}
\left(kv\right)^3+\tfrac{3\kappa^2}{4}kv =\tfrac{3\varepsilon
NU_0^2\eta_+^2 }\kappa t.\label{Eq03}
\end{equation}
The analytic solution of this cubic equation is shown as a solid
line in Fig.~2(c). In order to account for atomic debunching,
which is particularly strong during the one-sided switch-off
procedure, we have assumed that only $1/10$ of the atoms
participate in the CARL process. In contrast, the introduction of
friction to the dynamics leads to a steady state. At long times,
where $2kv\gg\kappa$, the differential equation in (2) is
approximately solved by
\begin{equation}
\left(  kv\right)^3=\tfrac{\varepsilon NU_0^2\eta_+^2}{\kappa
\gamma_{fric}}~.\label{Eq04}
\end{equation}

In summary, we studied the mutual backaction between the light
field of a unidirectionally pumped high-$Q$ ring cavity and cold
atoms trapped in this light field. For the first time, as far as
we know, strong evidence for the central role played by bunching
and by atomic recoil, has been found; the former by observing a
backscattered probe field in the absence of a seed and the latter
by detecting a displacement of the atomic sample. By introducing
an optical molasses the system approaches a steady state, which
might be interpreted as a \emph{cw} CARL. The atomic selfbunching
triggered by the molasses is an example for a phenomenon occurring
in extended dynamical systems: the enhancement or even the seeding
of spatio-temporal instabilities by a dissipative force.

\bigskip

We are grateful for helpful discussions with Helmut Ritsch. We
acknowledge financial support from the Landesstiftung
Baden-W\"urttemberg.

\end{document}